\documentclass[aps,twocolumn,superscriptaddress]{revtex4-1}
\usepackage{graphicx}
\usepackage{color}
\usepackage{ulem}
\begin{document}

\title{Charge Number Dependence of the Dephasing Rates of a Graphene Double Quantum Dot in Circuit QED architecture}

\author{Guang-Wei Deng}
\affiliation{Key Laboratory of Quantum Information, University of Science and Technology of China, Chinese Academy of Sciences, Hefei 230026, China}
\affiliation{Synergetic Innovation Center of Quantum Information \& Quantum Physics, University of Science and Technology of China, Hefei, Anhui 230026, China}
\author{Da Wei}
\affiliation{Key Laboratory of Quantum Information, University of Science and Technology of China, Chinese Academy of Sciences, Hefei 230026, China}
\affiliation{Synergetic Innovation Center of Quantum Information \& Quantum Physics, University of Science and Technology of China, Hefei, Anhui 230026, China}
\author{J.R. Johansson}
\affiliation{iTHES, RIKEN, Wako-shi, Saitama, 351-0198 Japan}
\author{Miao-Lei Zhang}
\affiliation{Key Laboratory of Quantum Information, University of Science and Technology of China, Chinese Academy of Sciences, Hefei 230026, China}
\affiliation{Synergetic Innovation Center of Quantum Information \& Quantum Physics, University of Science and Technology of China, Hefei, Anhui 230026, China}
\author{Shu-Xiao Li}
\affiliation{Key Laboratory of Quantum Information, University of Science and Technology of China, Chinese Academy of Sciences, Hefei 230026, China}
\affiliation{Synergetic Innovation Center of Quantum Information \& Quantum Physics, University of Science and Technology of China, Hefei, Anhui 230026, China}
\author{Hai-Ou Li}
\affiliation{Key Laboratory of Quantum Information, University of Science and Technology of China, Chinese Academy of Sciences, Hefei 230026, China}
\affiliation{Synergetic Innovation Center of Quantum Information \& Quantum Physics, University of Science and Technology of China, Hefei, Anhui 230026, China}
\author{Gang Cao}
\affiliation{Key Laboratory of Quantum Information, University of Science and Technology of China, Chinese Academy of Sciences, Hefei 230026, China}
\affiliation{Synergetic Innovation Center of Quantum Information \& Quantum Physics, University of Science and Technology of China, Hefei, Anhui 230026, China}
\author{Ming Xiao}
\affiliation{Key Laboratory of Quantum Information, University of Science and Technology of China, Chinese Academy of Sciences, Hefei 230026, China}
\affiliation{Synergetic Innovation Center of Quantum Information \& Quantum Physics, University of Science and Technology of China, Hefei, Anhui 230026, China}
\author{Tao Tu}
\affiliation{Key Laboratory of Quantum Information, University of Science and Technology of China, Chinese Academy of Sciences, Hefei 230026, China}
\affiliation{Synergetic Innovation Center of Quantum Information \& Quantum Physics, University of Science and Technology of China, Hefei, Anhui 230026, China}
\author{Guang-Can Guo}
\affiliation{Key Laboratory of Quantum Information, University of Science and Technology of China, Chinese Academy of Sciences, Hefei 230026, China}
\affiliation{Synergetic Innovation Center of Quantum Information \& Quantum Physics, University of Science and Technology of China, Hefei, Anhui 230026, China}
\author{Hong-Wen Jiang}
\affiliation{Department of Physics and Astronomy, University of California at Los Angeles, California 90095, USA}
\author{Franco Nori}
\affiliation{CEMS, RIKEN, Wako-shi, Saitama, 351-0198 Japan}
\affiliation{Physics Department, The University of Michigan, Ann Arbor, Michigan 48109-1040, USA}
\author{Guo-Ping Guo}
\email{Corresponding author: gpguo@ustc.edu.cn}
\affiliation{Key Laboratory of Quantum Information, University of Science and Technology of China, Chinese Academy of Sciences, Hefei 230026, China}
\affiliation{Synergetic Innovation Center of Quantum Information \& Quantum Physics, University of Science and Technology of China, Hefei, Anhui 230026, China}

\date{\today}

\begin{abstract}
We use an on-chip superconducting resonator as a sensitive meter to probe the properties of graphene double quantum dots (DQDs) at microwave frequencies. Specifically, we investigate the charge dephasing rates in a circuit quantum electrodynamics (cQED) architecture. The dephasing rates strongly depend on the number of charges in the dots, and the variation has a period of four charges, over an extended range of charge numbers. Although the exact mechanism of this four-fold periodicity in dephasing rates is an open problem, our observations hint at the four-fold degeneracy expected in graphene from its spin and valley degrees of freedom.
\end{abstract}
\maketitle

\begin{figure}[]
\includegraphics[width=\columnwidth]{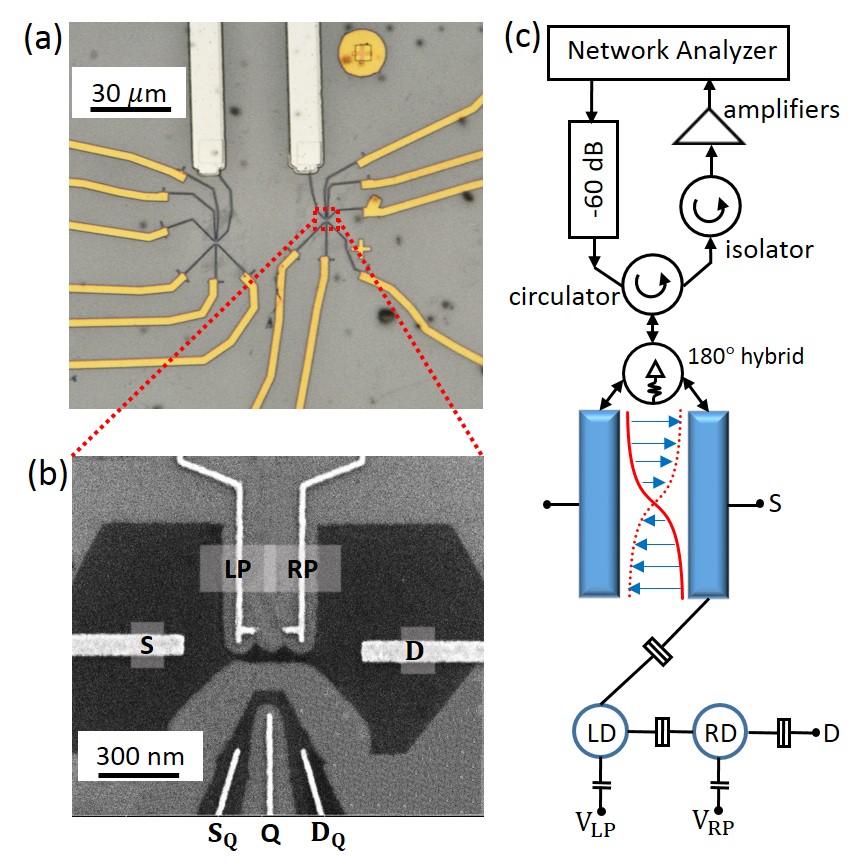}
\caption{ (color online). (a) Micrograph of the DQD gate structure. (b) Sample structure of a typical etched graphene DQD. The dc voltages used to control the charge numbers in the DQD are applied via left and right plunger (LP and RP) gates. A quantum point contact with a source ($S_Q$) and drain ($D_Q$) channel and a tuning gate ($Q$) is integrated near the DQD. (c) Circuit schematic of the hybrid device. The half-wavelength reflection line resonator is connected to DQD's left dot (LD) at one end of its two striplines. The right dot (RD) is connected to the drain. A microwave signal is applied to the other end of the resonator, and the reflected signal is detected using a network analyzer.}
\end{figure}

In recent years, on-chip microwave resonators have emerged as a useful tool both for coupling distant qubits and for sensitive metrology \cite{Xiang:RMP}. For example, many experimental studies have recently been performed to study the interaction between quantum dots (QDs) and resonators, in gate-defined carbon-nanotubes \cite{Delbecq:PRL,Delbecq:NC,Viennot:PRB}, GaAs \cite{Frey:PRL,Toida:PRL} and InAs nanowire \cite{Petersson:Nature,Liu:PRL} structures. Such studies are motivated by considering QDs as promising candidates for quantum information processing. Graphene has also attracted considerable attention in recent years because of its interesting physical properties and potential applications \cite{CastroNeto:RMP}. Like semiconductors, graphene-based QDs have been proposed as potential quantum bits \cite{Trauzettel:NP}. Various experiments are now underway to study the coherence properties of graphene QDs. For example, using pulsed-gate transient spectroscopy, Volk et al.~\cite{Volk:NC} measured a charge relaxation time of 100 ns in a graphene QD device. However, the dephasing times of grapheme QDs, which benchmark their quantum coherence and may be significantly shorter than their relaxation times, have not yet been measured. In addition, graphene has both spin and valley degrees of freedom, similar to carbon nanotubes \cite{Liang:PRL,Jorgensen:NP} and Si-based QDs \cite{Floris:RMP}. Spin qubits formed by graphene QDs have been theoretically studied \cite{Trauzettel:NP}, and various valley-related phenomena such as shell filling in carbon nanotubes \cite{Liang:PRL} and valley splitting in silicon \cite{Goswami:NP} have been explored. However, there are no experimental reports on the effects of the four-fold degeneracy caused by the spin and valley degrees in graphene devices.

Here, we present an experimental study of a graphene DQD device, which can be considered as a charge qubit that contains a large number of well-defined charge states. Using the sensitive dispersive readout of a microwave resonator, we measured the charge-state dephasing rates of this DQD in an integrated graphene-resonator device. Applying a quantum model describing the hybrid system, we simultaneously extracted: the DQD-resonator coupling strength, the tunneling rate between each quantum dot, and the charge-state dephasing rates. This microwave spectroscopy overcomes the difficulties of conventional transport techniques and allows us to study DQD dynamics in a large parameter space. In these experiments, we found that the dephasing rates depend on the number of charges in the dots. The rates vary with a periodicity of four charges, over an extended range of charge numbers, a behavior similar to the charging energy periodicity of carbon nanotubes \cite{Liang:PRL,Jorgensen:NP} and Si-based \cite{Floris:RMP} QDs. Stimulated by recent observations \cite{Shi:PRL,Shi:NC,Kim:Nature} that the dephasing rate is reduced for spin-charge hybrid qubit states in SiGe DQD systems, we speculate that our obervation may be caused by the hybrid states of spin and valley degrees in graphene DQDs. But further studies are needed to clarify this open problem.

Figure 1 shows the graphene-DQD/superconducting resonator device studied here. We have designed and fabricated a half-wavelength reflection line resonator (RLR) \cite{Vladimir:science,Zhang:APL} consisting of two differential microstrip lines. Contrasting traditional transmission mode designs, this design does not require the ground plane, and its microwave field is mostly confined between the two superconducting lines, where each point along the lines has an electrical potential with opposite sign (180-degree phase shift). The basic structure of the DQD along with the adjacent quantum point contact channel (black region in Fig. 1(b)) was defined by plasma etching of an exfoliated graphene flake \cite{Da:SR}. The two arms of the resonator were separately connected to the sources (S)of two DQDs [see Fig. 1(b,c)]. We only consider one of the DQDs in this experiment and all gates of the other DQD were always grounded. The samples were mounted in a dry dilution refrigerator with a base temperature of 26 mK. As the back gate is zero biased, the charge carrier in the DQD should be hole, according to our previous experiments \cite{Da:SR}. The resonator was coupled to a semi-rigid microwave transmission line via a 180-degree hybrid, which split the microwave signal into two components of opposite phases [Fig. 1(c)]. The reflected microwave signal was measured using a network analyzer. More details about the measurement setup can be seen in the supplementary materials \cite{Suppl:2014,pozar2004,Childress:PRA,Wiel:RMP,Collett:PRA,Kimble:PRA,Carmichael:PRL,Marco:PRB,Deng:arxiv}.

Figure 2(a) shows the transport current as a function of the two plunger-gate voltages. The two bright dots represent the triple points of the DQD, i.e., the only points where current flow is allowed because of the Coulomb blockade. We probed the DQD using the reflection line resonator by applying a coherent microwave signal to the resonator and then analyzing the reflected signal. Measuring the reflected microwave signal as a function of driving frequency, we determined the resonance frequency of the resonator to be 6.35076 GHz. With all the gates of the DQD grounded, the quality factor was $\sim$3000. We fixed the probe frequency at the resonant frequency and recorded the amplitude $A$ and the phase $\phi$ of the reflected coefficient $S_{11}$, as functions of the DQD gate voltages $V_{\rm LP}$ and $V_{\rm LP}$. The phase shift $\Delta\phi$ and amplitude change $\Delta A$ were most obvious at the triple points and on the inter-dot transition lines, where the charge states of the left and right dots become degenerate [see Fig. 2(b,c)]. The $\Delta\phi$ and $\Delta A$ are maximized at the detuning point. On other edges of the honeycomb, we found smaller values of phase shift and amplitude change, shown as faint lines. They were caused by the admittance change of the device when a charge state is changed \cite{Frey:PRB}. For this study we are only interested in the inter-dot configuration change, which occurs far from these charging lines and is dispersive in nature. We plot the amplitude change and phase shift as a functions of frequency in Fig. 2(d,e), for points B and A of Fig. 2(b), respectively. These full microwave spectra show that the changes originate from a shift in the resonance frequency of the reflected spectrum. Resonator frequency shifts are used to demonstrate the dispersive coupling between a charge qubit and a transmission resonator \cite{Frey:PRL}. Meanwhile phase shifts are used to demonstrate both charge and spin states coupling to a resonator \cite{Petersson:Nature}. In practice, if the probe frequency $\omega_{\rm R}/2\pi$ is kept fixed, while the cavity resonance frequency $\omega_{\rm 0}/2\pi$ decreases, both $\Delta A$ and $\Delta\phi$ are obtained \cite{Suppl:2014}.

\begin{figure}[]
\includegraphics[width=\columnwidth]{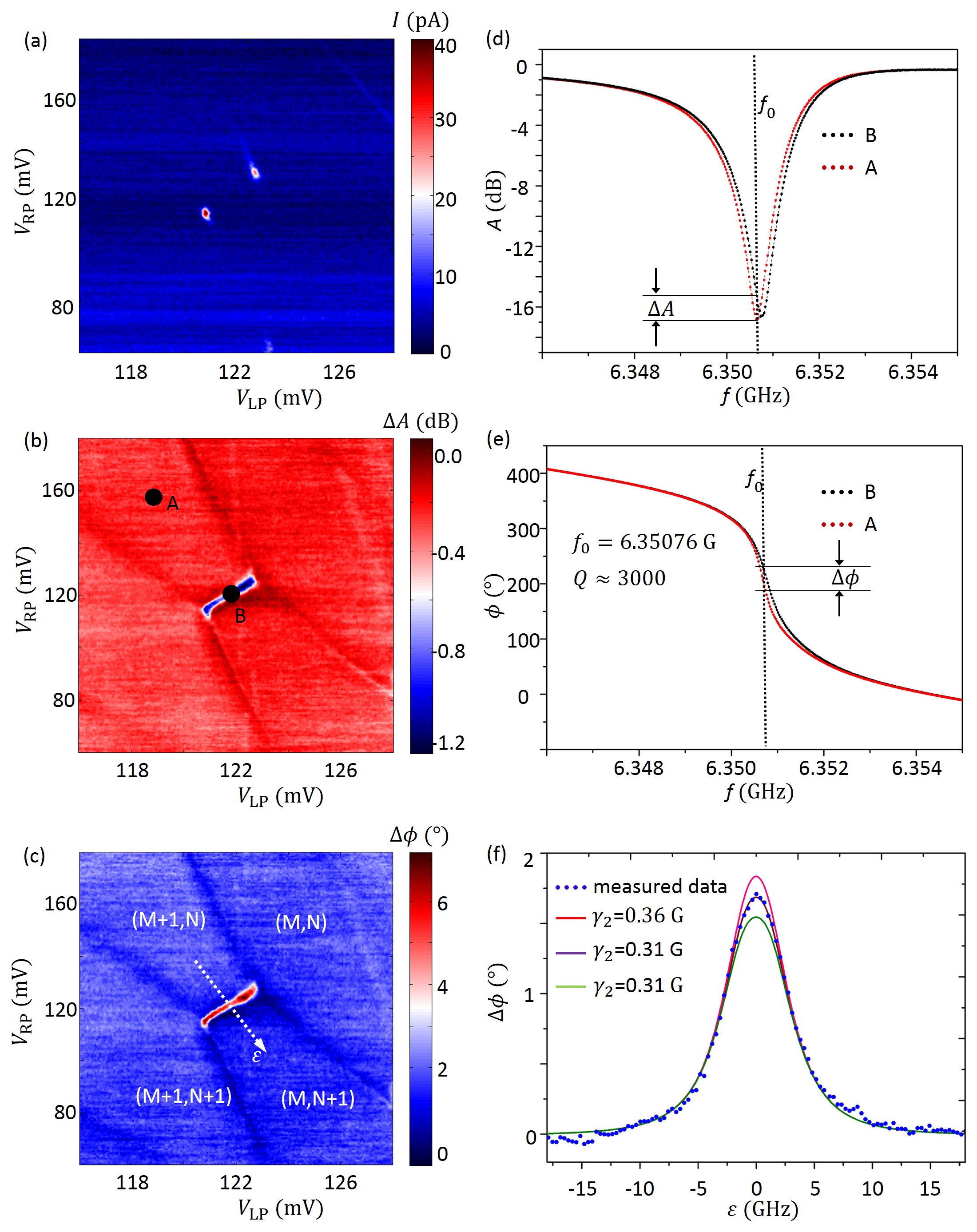}
\caption{ (color online). (a) The charge stability diagram measured using transport. (b-c) The charge stability diagram measured by the amplitude (b) and phase (c) response of the reflection line resonator. The three charge stability diagrams show a direct correspondence. (d) Amplitude response of the resonator, corresponding to point A and B in (b). (e) Amplitude response of the resonator. The resonance frequency and quality factor ($Q$) of the resonator can be extracted. (f) Fit of the phase response as a function of the detuning, thus $2t_{\rm C}$, $g_{\rm C}$ and $\gamma_2$ can be obtained.
}
\end{figure}

Having demonstrated good DQD-resonator coupling, we then tried to extract some of graphene's peculiar properties. For this hybrid system, the measured phase shift $\Delta\phi=-{\rm arg}(S_{11})$ depends on the resonance frequency $\omega_{\rm 0}$, driving frequency  $\omega$, internal and external resonator dissipation rates $\kappa_i$ and $\kappa_e$, DQD-resonator coupling strength $g_{\rm C}$, DQD interdot tunneling rate $2t_{\rm C}$, detuning energy $\epsilon$, relaxation rate $\gamma_1$, and dephasing rate $\gamma_2$.
We have used a quantum model to describe the graphene DQD and resonator hybrid system \cite{Suppl:2014}. Equipped with such model, here we investigate the inter-dot charge transition $(M+1;N)$ to $(M;N+1)$ by sweeping gate voltages along the detuning line, that is, across the corresponding transition line, while recording the phase and amplitude response \cite{Suppl:2014}. The reflection coefficient can be expressed as:
\begin{eqnarray}
S_{11}&=& -\frac{i(\omega_0-\omega)+g_{\rm eff}\chi+\frac{\kappa_{\rm i}-\kappa_{\rm e}}{2}}{i(\omega_0-\omega)+g_{\rm eff}\chi+\frac{\kappa_{\rm i}+\kappa_{\rm e}}{2}},
\end{eqnarray}
where $\chi=\frac{g_{\rm eff}}{i(\Omega-\omega)+\frac{1}{2}\gamma_1+\gamma_2}$, $\Omega=\sqrt{(2t_{\rm C})^2+\epsilon^2}$, $g_{\rm eff}=g_{\rm C}\frac{2t_{\rm C}}{\Omega}$. Here, $\omega_{\rm 0}$, $\kappa_i$ and $\kappa_e$ can be obtained by fitting the phase response as a function of probe frequency \cite{Suppl:2014}. Using temperature-dependent measurements and calculating with a charge network model, we can extract $g_{\rm C}$, $2t_{\rm C}$, and the lever arms of gates, which altogether give us a calibrated detuning $\epsilon$ \cite{Da:SR,Dicarlo:PRL}. In graphene QDs $\gamma_1$ has been reported to be less than 100 MHz \cite{Volk:NC}. Thus, the only unknown parameter is $\gamma_2$, which is both critical and unexplored for graphene QDs. In practice, by fitting the phase shift as a function of $\epsilon$, $g_{\rm C}$, $2t_{\rm C}$ and $\gamma_2$ can be obtained simultaneously. Figure 2(f) shows an example of extracting $\gamma_2$ by such fitting. We found $\gamma_2=0.31\pm0.02$ GHz for the charge state near the region $V_{\rm LP}$ = 123 mV and $V_{\rm RP}$ = 120 mV. The phase shifts at two other $\gamma_2$ values $\pm0.05$ GHz away from the optimal value are clearly different from that at the optimal value, which demonstrates the sensitivity of this parameter extraction method. This method has also been extensively used in other studies of the circuit QED of GaAs \cite{Frey:PRL,Toida:PRL}, InAs nanowire \cite{Petersson:Nature} and carbon nano-tube \cite{Delbecq:PRL} DQDs. We systematically extracted $\gamma_2$ for different charging states, observing dephasing rates over the range 0.3 GHz to 10 GHz [see Fig. 3(b)]. The lowest dephasing rate in our graphene DQDs represents a lower bound of the dephasing rate, caused by charge fluctuations in the environment. The dephasing rate of 0.3 GHz is comparable to that in GaAs \cite{Frey:PRL,Toida:PRL} and carbon nanotube \cite{Viennot:PRB} DQDs. Because the dephasing rate in grapheme DQDs has not been obtained by any other means namely photon-assisted tunneling (PAT), a traditional method, we now speculate here a possible reason. Using traditional methods involving charge transport, determining $\gamma_2$ is easily masked by the puddle and edge states \cite{Evaldsson:PRB,Gallagher:PRB,Wang:APL} in an etched graphene structure. However, the resonant cavity here is mostly sensitive to the electrical dipole of the DQD and is affected much less by electrostatic disorder in the etched grapheme structure.

Graphene has both spin and valley degrees of freedom, and graphene DQDs can have complicated energy levels in many-charge regions. However, it is difficult to control and identify the energy structure of graphene QDs by traditional transport methods such as PAT, possibly because of edge states and puddles \cite{Evaldsson:PRB,Gallagher:PRB,Wang:APL}. The sensitive microwave-based metrology used here allows us to study the dephasing rates in graphene DQDs, and whether they depend on the charge number in a DQD. This is motivated by recent reports, which have noted that, in SiGe DQDs, the coherent time of the (2,1) charge state can be much longer than the (1,1) state, because its energy dispersion depends on the spin degree of freedom \cite{Shi:PRL,Shi:NC,Kim:Nature}. This finding gives meaningful insight into using different charge-number states to encode a qubit.

We found a large charge stability diagram in our device, which contains many well-shaped honeycomb patterns [see Fig. 3(a)]. This diagram lets us study how the dephasing rates vary with charge number over an extended range, revealing that the rate varied considerably with charge number. As shown in Fig. 3(b), the rate varies by as much as a factor of $\sim$30. Such large variation is unexpected. It is in direct contrast with the behavior observed in semiconductor QDs, where Basset et al. \cite{Basset:PRB} found that the dephasing rate in a GaAs DQD varies little between the few-electron and many-electron regimes, which suggests that charge number had only minor effect in that system.

An intriguing observation in our experiment is that the dephasing rate not only varied significantly, but the pattern of the variation also appeared to repeat every four charges for the right dot. This is clearly shown in Fig.~3(b) for 12 consecutive rows $(n, n+1, ... ,n+11)$ of honeycomb cells. We have verified this pattern for five consecutive columns $(m, m+1, ... ,m+4)$. To our knowledge, such periodic variation of a physical quantity has not been seen in any experiment on graphene QDs. Basset et al. \cite{Basset:PRB} found that the dephasing rate depends on $2t_{\rm C}$ for a fixed charge state; however, $2t_{\rm C}$ in our device is tuned to be around 6-8 GHz \cite{Suppl:2014}, which should not be seen as a main source of the observed periodicity. While the data fitting process may have an $\sim20\%$ error, which is mainly caused by the transformation of $\epsilon$ \cite{Da:SR}, the periodic variation is not affected. The relaxation rate is fixed at 100 MHz in the fitting process. Also, any potential sub-GHz-order variation in the relaxation rate would not have changed the GHz-order periodicity in the total decoherence rate. In addition, we repeated these experiments in other two similar samples with the same sample structure, producing similar results. To be consistent, here we only present data from one sample, named sample A. Results of other two samples, named sample B and C, are shown in the supplementary. While the four-fold periodicity is clear when the charge number in the right dot was varied, it is less conclusive when the charge number in the left dot was varied because there were only five ordered columns in our experiments \cite{Suppl:2014} [see Fig.~3(a)].
\begin{figure}[]
\includegraphics[width=\columnwidth]{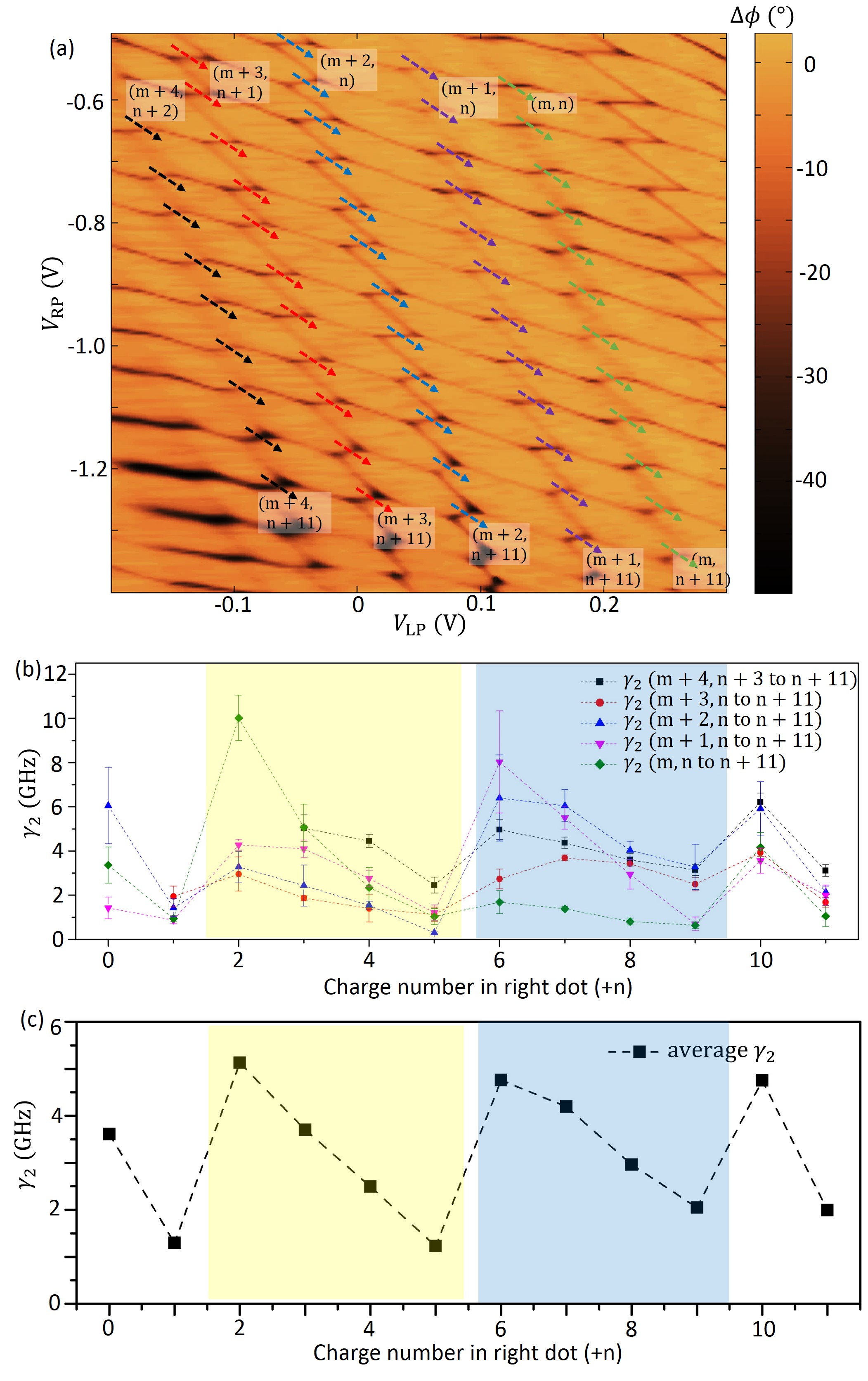}
\caption{ (color online). (a) Charge stability diagram of device A. The diagram contains about $5\times12$ honeycomb cells, which indicates the region for our charge-number-dependence study, since the shape of the honeycombs is well-defined and the resonator signal is sufficiently strong in this region. The charge numbers are indicated for both the left and right dots. (b) The dephasing rates as a function of hole number in the right dot for 12 consecutive holes (from $n$ to $n+11$). Five columns (from $m$ to $m+4$) are studied. To assist readers, the approximate periodicity has been guided by different background colors. (c) The averaged dephasing rates of (b) shows a periodicity.
}
\end{figure}

We believe that this periodicity in the charge dephasing rate could be an indication of the four-fold valley-spin degeneracy in a graphene QD. In past studies of carbon nano-tube DQDs, such degeneracy has manifested itself in a variation of the charging energy \cite{Jorgensen:NP}. However, periodic variations of the charging energy are not observed in our system \cite{Suppl:2014}. As is well known, if the single-particle excitation energy is much smaller than the charging energy, any periodic variation will likely be masked by measurement noise \cite{Liang:PRL}. Although the energy splitting due to different spin and valley occupation cannot be resolved by transport measurements, the energy dispersion of the DQD can definitely be changed by the spin and valley coupling \cite{Shi:PRL}. Because the charge noise is very sensitive to the microscopic details of the energy spectrum, such as the bias dependence of the energy levels, the dephasing rate, given by $\frac{d^2\Omega}{d\epsilon^2}\mid_{\epsilon=0}\langle\sigma_\epsilon\rangle$ (Here $\langle\sigma_\epsilon\rangle$ is the noise term) \cite{Viennot:PRB,Basset:APL} can vary periodically as the spin and valley quantum numbers are periodically altered.

A variation of the dephasing rate caused by energy dispersion has been observed in a Si/SiGe DQD \cite{Shi:PRL,Shi:NC}. In a graphene DQD, many parameters are still unknown, such as the exact DQD charge states, as well as the spin and valley splitting energies. In our measurement there is a clear trend that $\gamma_2$ decreases monotonously with charge number in a filling period. It seems that a partially-filled valley-spin shell provides some screening to smooth the energy dispersion of our DQD and enhances the coherence times \cite{Shi:PRL,Shi:NC,Kim:Nature}. Clearly, a comprehensive understanding of our observation here requires a detailed theoretical analysis of the electronic structure of a multi-hole double dot in graphene, which is beyond the scope of this paper. We believe that this result, if confirmed, could be an important clue for a better optimized QD-based qubit design. We thus hope that our finding stimulate further theoretical and experimental studies that give a better microscopic understanding of this interesting phenomenon.

In summary, we have implemented a half-wavelength reflection line resonator, and coupled the resonator to a graphene DQD. This platform let us study the physics of light-matter interaction with graphene devices in the microwave regime. In this hybrid device we demonstrated a graphene DQD-resonator coupling strength in the tens of MHz. By fitting the phase shift as a function of the DQD detuning, we characterized the device and extracted the charge dephasing rates of the system by using a quantum model. The many well-shaped honeycomb cells in our device allowed us to observe the 4-electron periodicity of the dephasing rates, which may hint at the four-fold degeneracy expected for the twice two-fold valley and spin level degeneracy in graphene.

This work was supported by the National Fundamental Research Programme (Grant No.~2011CBA00200), and National Natural Science Foundation (Grant Nos.~11222438, 10934006, 11274294, 11074243, 11174267 and 91121014).

\bibliography{ref}

\end{document}